# Adsorption-controlled Growth of Homoepitaxial *c*-plane Sapphire Films

Lena N. Majer,[1] Tolga Acartürk,[1] Peter A. van Aken,[1] Wolfgang Braun,[1] Luca Camuti,[1,2] Johan Eckl-Haese,[1] Jochen Mannhart,[1] Takeyoshi Onuma,[3] Ksenia S. Rabinovich,[1] Darrell G. Schlom,[4,5,6] Sander Smink,[1] Ulrich Starke,[1] Jacob Steele,[4] Patrick Vogt,[1,7] Hongguang Wang,[1] Felix V.E. Hensling[*,1,4]

[1] Max Planck Institute for Solid State Research, 70569 Stuttgart, Germany

[2] Ludwig-Maximilians-Universität, Department of Chemistry, 80539 Munich, Germany

[3] Department of Applied Physics, Kogakuin University, 2665-1 Hachioji, Tokyo 192-0015, Japan

[4] Department of Materials Science and Engineering, Cornell University, Ithaca, NY14853, USA

[5] Kavli Institute at Cornell for Nanoscale Science, Ithaca, NY 14853, USA

[6] Leibniz-Institut für Kristallzüchtung, 12849 Berlin, Germany

[7] Institut für Festkörperphysik, Universität Bremen, 28359 Bremen, Germany

[*] f.hensling@fkf.mpg.de

Sapphire is a technologically highly relevant material, but it poses many challenges to performing epitaxial thin-film deposition. We have identified and applied the conditions for adsorption-controlled homoepitaxial growth of *c*-plane sapphire. The films thus grown are atomically smooth, have a controlled termination, and are of outstanding crystallinity. Their chemical purity exceeds that of the substrates. The films exhibit exceptional optical properties such as a single-crystal-like bandgap and a low density of $F^+$ centers.

# 1. Introduction

Sapphire (α-Al$_2$O$_3$) is a material that features highly valuable properties for fundamental research and applications. It is especially useful as a substrate and functional film for electronic applications. Sapphire combines a large band gap (9 eV), high dielectric permittivity (9–11), low RF losses, high thermal conductivity (46 W/mK), and a high electric breakdown field with remarkable thermal and mechanical stability [1–4]. The use of sapphire as a substrate material has been particularly driven by the ability to deposit high-quality silicon on sapphire for field effect transistor devices [5] and to fabricate GaN-based light-emitting diodes on sapphire [6]. As a result of these developments and owing to the availability of low-cost, high-quality, single-crystal sapphire wafers of practically any size, the ever-growing market share of sapphire substrates has already surpassed that of any other oxide substrate material [1]. Furthermore, being a functional material itself, sapphire is a frequently used high-$k$ gate dielectric [7] and is utilized as an active material in solid-state lasers [8,9]. Sapphire has also gained attention as an ultrawide-band-gap material for high-power electronics [10–12].

However, the growth of high-quality sapphire films poses several challenges for further important applications of sapphire in electronics. High-quality sapphire films are required for technologies that rely on active or passive sapphire films and, notably, for the use of sapphire as a substrate material. As substrate surfaces are usually rich in impurities and defects induced, for example, by the mechanical processing of the substrates, epitaxial layers of the substrate material are preferentially deposited on the substrate to shield devices from likely blemished substrate surfaces [13–15]. Whereas the growth of sapphire single crystals has long been refined, the epitaxial deposition of high-quality sapphire films has proved to be difficult. This is in part because aluminum effusion cells are notoriously difficult to handle in the oxidizing environments needed for sapphire growth by molecular beam epitaxy (MBE). For example, the tendency of Al to creep requires the use of cold-lip effusion cells. However, aluminum oxide tends to form at their orifices and clog them, thus leaving the effusion cell useless after short times of usage [16]. Nevertheless, *a*-, *m*-, and *r*-plane-oriented sapphire films have been grown by MBE, although they lack atomically flat surfaces, [11,12,17,18] and atomically smooth *r*-plane films have been realized by pulsed laser deposition (PLD).[19] Furthermore, compared to single crystals, the optical band gap of these films is reduced, a degradation that has been attributed to impurities embedded in the films partially originating from the effusion cell crucible [11,20]. Even worse, owing to the formation of the preferred γ-phase of $Al_2O_3$, the epitaxial growth of the highly important *c*-plane sapphire films is not even considered possible with MBE [12,18,21]. Driven by the technological relevance of the *c*-plane sapphire surfaces, several efforts have been undertaken to resolve this impediment. Epitaxial *c*-plane sapphire films, for example, have been grown by electron-beam-assisted PLD. However, the films thus achievable were only several unit cells thick and partially amorphous [22].

In order to provide high-quality sapphire films, we explored the growth of *c*-plane sapphire by adsorption control. Adsorption-controlled growth implements the self-limiting, stoichiometric, epitaxial growth of thin films. For film growth by adsorption control, the deposition parameters, i.e., substrate temperature and gas supply, are chosen such that the desired phase of the film to be grown is stable, but that all except one of the molecular species arriving at the growing film are volatile. Therefore, when growing stoichiometric films, these species may be supplied in larger fluxes than required to match the desired film composition one-to-one because the excess molecules desorb from the growing film. Adsorption-controlled growth, used with stunning success for depositing III–V heterostructures [23–26], is typically the growth mode that yields the cleanest films of the highest crystalline quality [27,28]. However, adsorption-controlled growth is limited to the growth of materials for which the required deposition parameters can be achieved.[29] It remains to be explored whether the benefits of adsorption control outweigh the thermodynamic drive of an increased point defect formation at the potentially high temperatures required. The growth of oxides by standard adsorption control has been mostly limited to oxides containing highly volatile species, such as ruthenates [30–32], iridates[33–35], bismuthates [36–38], plumbates [39–41] and stannates [42–44]. To overcome the limitation to high volatility, organic precursors have also been used, for example to grow SrTiO$_3$ [45–47], although these may introduce undesirable impurities at low substrate temperatures [48]. Films of group-III oxides have been grown by adsorption control using suboxide-MBE. Here, the typically high volatility of the suboxide formed in the first reaction step of the oxidation of group-III elements (Eq. (1)) is utilized, with it being supplied in excess [15,49,50]. However, this method is more difficult to apply to sapphire films, as both Al and its suboxide are not as volatile [51]. For this reason, adsorption-controlled growth of sapphire has long been a highly desirable epitaxial growth process, albeit one that has yet to be realized.

$$2\text{III}(g) + 3\text{O}(g) \overset{1\text{st}}{\leftrightarrow} \text{III}_2\text{O}(g) + 2\text{O}(g) \overset{2\text{nd}}{\leftrightarrow} \text{III}_2\text{O}_3(s). \tag{1}$$

In this work, we explore and reveal the parameter space necessary for the adsorption-controlled growth of sapphire. We find these parameters to be barely achievable with standard MBE. In comparison, the extended parameter space of thermal laser epitaxy (TLE) [29,52] allows us to grow atomically smooth films at high growth rates. The crystallinity and band gap of these films are bulk-like: the film purity exceeds that of the substrate. Epitaxial growth is found to be possible up to an effective substrate temperature of 2000 °C, the temperature at which the back side of the substrate facing the heating laser beam already melts, while films can still be grown on the front side of the substrate. The two key ingredients for achieving high quality ultra-pure sapphire films are thus high substrate temperatures and avoiding impurities from parasitic heat and hot crucibles.

## 2. Methods

*Epitaxial growth and growth analysis*: All CrysTec GmbH *c*-plane sapphire substrates (miscut <0.1°) were annealed in vacuum at a background pressure of <1×10$^{-7}$ mbar at 1700 °C for 200 s prior to growth. This results in a pure Al-rich surface termination with a double layer step height as shown in Fig. 1a-b [53]. MBE films were grown in a Veeco Gen10 and thermal laser epitaxy (TLE) films in an epiray STRATOLAS 50 with a THERMALAS heater. 5N aluminum was loaded in an p-BN crucible in a mid-temp effusion cell for MBE growth and in a sapphire crucible for TLE growth. For MBE this means that over time a clogging of the effusion cell is expected. Thus, the flux needs to be monitored closely – an issue not present in TLE as the source laser simply breaks through any oxide crust forming. The TLE Al source is operated in a regime, where an Al flux with a neglectable $Al_2O$ portion is expected [16]. The oxygen source for MBE supplied a mixture of ≈ 80% $O_3$ with $O_2$, the one for TLE 100% $O_2$ (purified). It can be considered that $O_3$ dissociates to release O at the growth front.[54] Substrate temperatures in both systems were measured by a pyrometer. Film growth was monitored *in situ* by reflection high-energy electron diffraction (RHEED) before and after growth because monitoring during growth produces surface defects. A Bruker DektakXT depth profilometer was utilized to measure the thickness of the resulting films along the position of the sample holder during growth (compare Fig. 2c). To exclude any other effects resulting in shadowing by the substrate holder, in a control experiment a sapphire substrate was heated to 2000 °C without deposition. In this case no height difference was observed, as expected. An Asylum Cypher atomic force microscope (AFM) in tapping mode was used to acquire the topography images and a Panalytical Empyrean for X-ray diffraction (XRD).

Considering the isobaric-isothermal nature of the sapphire growth in TLE, *ΔG(T)* is given by

$$\Delta G(T) = \Delta H(T) - T\Delta S(T) \tag{2}$$

with the change in enthalpy (*ΔH(T)*) and the change in entropy (*ΔS(T)*) determined by

$$\Delta H(T) = \Delta H_0 + \int_{T_0}^{T} dT\, C(T) \quad \text{and} \tag{3}$$

$$\Delta S(T) = \Delta S_0 + \int_{T_0}^{T} dT\, \frac{C(T)}{T} \tag{4}$$

with $T_0$ = 295 K. The heat capacity *C(T)* is calculated as

$$C(T) = a + b10^{-3}T + c10^{6}T^{-2} + d10^{-6}T^2. \tag{5}$$

$\Delta H_0$, $\Delta S_0$, $a$, $b$, $c$, and $d$ are adapted from Ref. [55]. For a given chemical reaction with reactants $R_i$ and products $P_j$, $\Delta G < 0$ can be determined by the sum of the Gibbs free energies of the products, $\Sigma_j G_{P_j}$, minus the sum of the Gibbs free energies of the reactants, $\Sigma_i G_{R_i}$, i.e., by

$$\Delta G = \sum_j p_j G_{P_j} - \sum_i r_i G_{R_i}, \tag{6}$$

with stoichiometric coefficients $p_j$ and $r_i$. The plot in Fig. 3d is derived from Eqs. (2) – (6).

*Transmission electron microscopy:* TEM specimens of the films were prepared by mechanical wedge polishing followed by Ar ion beam milling at L-N$_2$ temperature (Micron 140 (2021) 102979). Scanning transmission electron microscopy (STEM) studies were performed using a spherical aberration-corrected STEM (JEM-ARM200F, JEOL Co. Ltd.) equipped with a cold-field emission gun and a DCOR probe Cs corrector (CEOS GmbH) operated at 200 kV. The STEM images were obtained using an ADF detector with a convergent semi-angle of 20.4 mrad. The corresponding collection semi-angles for high-angle annular dark-field (HAADF) and low-angle ADF (LAADF) imaging were 70–300 mrad and 40–100 mrad, respectively. Eight serial images were acquired with a short dwell time (2 µs/pixel), aligned, and then added to improve the signal-to-noise ratio (SNR) and to minimize distortion of the HAADF and LAADF images. STEM images were denoised using the band-pass filter.

*ToF SIMS:* ToF SIMS was measured using a ToF.SIMS NCS instrument (IONTOF). For mass-spectrometry probing, monoatomic Bi$^{1+}$ ions were used in spectrometry mode, accelerated by a voltage of 30 keV, with currents in the range of 2.1 to 2.4 pA on a 50 × 50 µm² analysis area. Ion spectra were acquired with positive and negative polarity to access a variety of different atomic species. Data analysis was performed using Surfacelab 7.1 software. Depth profiles were acquired by a cyclic sputter-probe series. (Note that only data for positive ions are shown in the paper.) The sputter and analysis guns were operated in non-interlaced mode with an additional flood gun in order to avoid sample charging. For sputtering, an oxygen ion source was employed using 2 keV acceleration voltage and 479 nA current on a 400 × 400 µm² crater area. Ion intensities shown in the depth profiles are averaged 10-fold.

*Optical properties*: The real part of optical conductivity $\sigma_1$ was determined from ellipsometric parameters $\Psi$ and $\Delta$ measured by a variable-angle spectroscopic ellipsometer VASE (J. A. Woollam Inc.) in the energy range from 1.5 to 6.0 eV. $\Psi$ and $\Delta$ are defined by the complex Fresnel coefficients as $\tan(\Psi)e^{i\Delta}=r_p/r_s$, where $r_p$ and $r_s$ are reflection coefficients for light polarized parallel and perpendicular to the plane of incidence. The optical measurements were performed at room temperature and at various angles of incidence (from 70° to 80°). UV-Vis spectra were obtained utilizing a Cary 5000 (Agilent Technologies) capable of measuring absorptance in an integrating sphere. Transmittance spectra in UVC spectral range were measured at room temperature (RT) using a monochromatic light in a wavelength range of 135–300 nm from a deuterium lamp (Hamamatsu L11798) dispersed by a 20-cm focal-length Czerny–Turner monochromator (Bunkoukeiki KV-200) equipped with a 1200 groove/mm grating [56]. The samples were set on the rotational sample stage, which was loaded into the vacuum chamber. The mirror optical system and monochromator were connected to the chamber through $MgF_2$ viewports. During the measurements, the entire optical path was purged with ≥99.99% pure nitrogen gas with a gas flow rate of 7 L/min to maintain a residual oxygen concentration lower than 1 ppm. Spectral resolution at 160 nm was set at 0.1 eV (2 nm) with a slit width of 0.5 mm. A 0.5 × 0.6 mm² rectangular light spot was defined by the slit widths of the monochromator and the luminance point size of the $D_2$ lamp. The transmitted light was detected using a photomultiplier tube (Hamamatsu R374) through a window coated with sodium salicylate. A quartz glass long-pass filter was used in the wavelength range greater than 200 nm to exclude the second-order light. Lock-in detection using an optical chopper was employed for the measurements.

## 3. Results & Discussion

Attempts at homoepitaxial growth of *c*-plane sapphire usually results in the formation of the $Al_2O_3$ γ-phase which, not being sapphire, is undesirable for electronic applications [12,18]. Likewise, reflection high-energy electron diffraction (RHEED) patterns reveal amorphous growth ([Figs. 1c, S1](Figs. 1c, S1)) for both MBE and TLE growth at substrate temperatures $T$ < 900 °C and the parameters found in Tables 1 and 2. The surfaces of these films are rough with a root-mean-square (RMS) value >5 nm (Figs. 1d, S1). The difficulty of obtaining the α-phase (sapphire) is determined by the high temperature required for its phase formation under equilibrium conditions [57]. Consequentially, growth at higher $T$ (900 °C ≤ $T$ < 1000 °C) yields crystalline homoepitaxial films for MBE and TLE, as confirmed by streaky and spotty RHEED patterns (Figs. 1e, S1). Nevertheless, the films remain rough (Figs. 1f, S1). As in most MBE systems with non-laser substrate heaters, the $T$ achievable with the MBE setup we used is limited.[29] Although a homoepitaxial deposition of sapphire films is, thus, achievable under the conditions described, their surfaces are not smooth.

Enabled by a $CO_2$-laser-based substrate heating system, substrate temperatures >>1000 °C are available in the TLE system used here.[29,52] Indeed, deposition at 1100 °C yields much smoother films with RMS ≈ 0.5 nm, see Fig. 1h. The islands visible in this figure follow the quasi-hexagonal motif of the crystal structure and are atomically smooth with one monolayer step heights. The respective RHEED pattern (Fig. 1g) shows a complex reconstruction not previously reported probably resulting from the island structure. Thus, in the multilayer growth mode the single-reconstruction surface of the prepared sapphire substrates is not preserved. The observed islands are likely a result of kinetic roughening as observed for the homoepitaxial growth of other materials.[58] The island size increases with increasing temperature until a shift towards the growth regime discussed below is observed.

Increasing $T$ beyond 1300°C results in a transition from the multilayer growth mode to step-flow growth. Independent of the film thickness, the resulting surfaces are practically indistinguishable from those of the prepared substrates. Figure 1j shows the corresponding surface of a film grown at 1600 °C (RMS = 0.23 nm). Each of the uniform surface steps, which border terraces with a remarkable width of ≈2 μm, has a height of two layers along [0001]. These double steps and the Al-terminated √31×√31R+9° reconstruction (Fig. 1i) are also present on the prepared substrates [53]. These films thus retain the Al-termination and atomic smoothness of the substrate.

This nominally ideal surface structure is observed up to 1800 °C. At higher $T$, step bunching occurs. This is illustrated in Fig. 2a, which shows a film surface after growth at $T \approx 2000$ °C. As mentioned above, the back side of the substrate melts at $T \approx 2000$ °C, generating the melt pool and bubbles visible in the optical micrograph. Nevertheless, the sapphire films grown on the slightly cooler front side show sizable step bunching (Fig. 2b). Figure 2b also shows the sample areas in which film growth is blocked by a shadow cast by the substrate holder such that the film thicknesses can be measured straightforwardly by a profilometer at these places. The steps reach such a height that they are visible to the naked eye, whereas the terraces remain atomically flat as shown by AFM with step heights of various multitudes of two layers along [0001] (Fig. 1l). The RHEED images still represent a singly oriented √31×√31R+9° reconstruction (Fig. 1k) showing that the Al-rich surface termination is stable on the entire sample even for growth at $T \approx 2000$ °C.

The different growth regimes observed above are for substrates of similar miscuts. Changes in miscut may alter the transition temperatures. Additionally, the surface structure of $c$-plane sapphire is miscut dependent.[59,60] We observed, however, that in the step-flow growth regime the initial surface structure was maintained.

Figures 1 and S1 reveal a drastic change at $T = 900$ °C where the growth for both MBE and TLE transitions from amorphous to epitaxial $c$-plane sapphire (summarized in Table 1). Therefore, we investigate whether this transition also affects the growth rate $\Gamma$. Indeed, the growth rate $\Gamma$ decreases at $T = 900$ °C for both MBE and TLE film growth (Fig. 3a), suggesting that the $c$-plane sapphire films grow in an adsorption-controlled manner for $T \geq 900$ °C up to a hypothetical upper limit (beyond the melting point of the sapphire substrate) – meaning that each species not forming sapphire is desorbed. To explore whether this is indeed the case, we recall the two reaction steps for the formation (Eq. 1) of sapphire by oxidation of Al: [49,50,61–63]

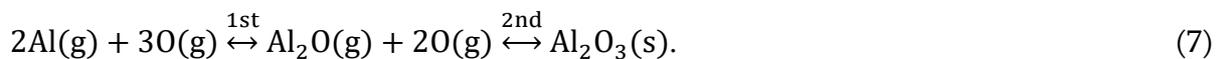

$$2Al(g) + 3O(g) \overset{1st}{\leftrightarrow} Al_2O(g) + 2O(g) \overset{2nd}{\leftrightarrow} Al_2O_3(s). \qquad (7)$$

We consider O to result from the O3 dissociation and thermalization of O2, respectively. In reality the same reaction with $O_2$ as a reactant may occur. Thus, in the hypothetical parameter range of adsorption-controlled growth, $Al_2O$ must be volatile (additionally Al can be volatile as well). This $Al_2O$ volatility is expected to have three effects on the $Al_2O_3$ growth rate $\Gamma$: [49,50]

(1) $\Gamma$ tends to be small compared to standard growth because $Al_2O$ desorbs.

(2) $\Gamma$ will be independent of the Al supply because $\Gamma$ is limited only by the oxygen/ozone flux, and therefore

(3) $\Gamma$ will increase with the flux of oxygen/ozone impinging on the substrate.

For adsorption-controlled growth, these three effects are highly characteristic. We will use them as a fingerprint of sapphire film growth by adsorption control.

First, exploring the existence of effect (1), we find that, for the parameters listed in Table 2, the growth rate of sapphire films is smaller at $T \geq 900$ °C than at lower $T$, where the films do not exhibit the $\alpha$-phase, for MBE as well as for TLE (Fig. 3a). This behavior is in agreement with effect (1). Note that the $\Gamma$ values achievable for growth by MBE ($\approx 4$ μm/h) and by TLE ($\approx 15$ μm/h) at $T < 900$ °C are considerable. Even at $T \geq 900$ °C $\Gamma$ is high, namely $\approx 0.5$ μm/h.

Also, effect (2) is present, as shown by Fig. 3b, which displays the dependence of $\Gamma$ on the source laser power $P_L$ for $p_{ox} = 2 \times 10^{-3}$ mbar and $T = 1600$ °C. The source laser power correlates to the flux as described in detail in Refs. [64,65] and for Al sources specifically in Ref. [16]. For comparison, for the same $P_L$, Fig. 3b shows $\Gamma$ also for film deposition on a nominally unheated substrate ($T \approx 80$ °C). For the latter case, an Arrhenius-like dependence of $\Gamma$ on $P_L$ is expected [16], which is indeed observed. However, at $T = 1600$ °C and $p_{ox} = 2 \times 10^{-3}$ mbar, $\Gamma$ does not depend on $P_L$ and therefore not on the Al supply.

To discuss the existence of effect (3), we present the dependence of $\Gamma$ on $p_{ox}$ in Fig. 3c. For the unheated substrate, $p_{ox}$ does not alter $\Gamma$. For $p_{ox} > 9.6 \times 10^{-3}$ mbar, for which the mean free path is shorter than the substrate-source distance, $\Gamma$ is lower than at smaller $p_{ox}$. At $T$ = 1600 °C and for $1\times10^{-3}$ mbar $\leq p_{ox} \leq 9.6\times10^{-3}$ mbar $\Gamma$ is proportional to $p_{ox}$, in agreement with effect (3). $\Gamma$ vanishes for $p_{ox} < 1\times10^{-3}$ mbar. A likely explanation is the lack of oxygen to complete the second reaction step of Eq. (7). All available oxygen goes to form $Al_2O$, which is volatile at such values of $T$ and therefore desorbs. There is also no growth observed for even lower pressures, indicating that also elemental Al is volatile at these $T$. For $p_{ox} \geq 1\times10^{-3}$ mbar, however, $\Gamma$ increases with $p_{ox}$. With more oxygen available, the oxidation of $Al_2O$ to $Al_2O_3$ increases, thus reducing $Al_2O$ desorption.

For the parameters in Table 2, sapphire growth shows all three effects that together provide a fingerprint of the adsorption-controlled regime. Therefore, we conclude that the adsorption-controlled growth of $c$-plane sapphire is indeed achievable.

For the growth process, there may appear to be some discrepancy between the conclusion that the films grow in an adsorption-controlled manner and that the growth rate $\Gamma$ is approximately independent of $T$ for $T > 900$ °C, see Fig. 3a. In fact, however, there is no discrepancy.

At first glance, the $T$-independence of $\Gamma$ is unexpected because one would expect that $\Gamma$ decreases with increasing $T$ for the adsorption-controlled regime. This is because adsorption decreases with $T$, as is the case for many group-III oxides. However, a closer look at the Gibbs free energies of the reactions involved in film growth reveals that, for the growth of $Al_2O_3$ films, this saturation of $\Gamma$ may be caused by Al chemically decomposing the $Al_2O_3$ to $Al_2O$, see Eq. (8) and Fig. 3d. The increased $Al_2O$ formation may in turn have a self-catalytic effect on the growth, ultimately resulting in the more or less stable growth rate at $T > 900$ °C.

Figure 3d provides data for the Gibbs free energy differences for all expected reactions ($\Delta G$) as a function of $T$. It omits reactions that are not expected for kinetic limitations.[66–70] The data shown was calculated as described in the Methods section based on values published in Refs. [71,72]. It is apparent that, in addition to the two reaction steps of Eq. (7), the above-mentioned etching process

$$Al_2O_3(s) + 4Al(g) \leftrightarrow 3Al_2O(g) \tag{8}$$

exists with $\Delta G < 0$. Note that, as shown in Fig. 3d, $\Delta G$ becomes slightly more negative with increasing $T$. For this case, the standard model used to characterize the adsorption-controlled growth of group-III oxides [73] applied to Eqs. (7) and (8) readily yields a good fit to the measured data (Fig. 3a, dashed line). Despite this success, this fit must be regarded with skepticism because, for example, the underlying model does not consider the effects of the $T$-dependent adhesion and surface diffusion processes on $\Gamma$. Self-catalysis as an explanation for the observed growth plateau is at this point, thus, to be considered a hypothesis. Extended experimental work and models will show if it is indeed the underlying mechanism.

The data shown in Figs. 1 and 3 allow us to deduce the ideal growth regime for the adsorption-controlled homoepitaxial growth of $c$-plane sapphire, which is also summarized in Table 2. The range 1300 °C < $T$ ≤ 1800 °C results in atomically flat surfaces with a pure Al-termination under these conditions. Utilizing high oxygen background pressures allows us to maximize the growth rates, in this particular case to $\Gamma \approx$ 3 μm/h.

Having established the optimal parameters for the adsorption-controlled growth of $c$-plane sapphire, we now explore the properties of films grown at $T$ = 1600 °C and $p_{ox}$ = 8×10$^{-3}$ mbar, i.e. adsorption controlled.

The RHEED patterns of the films, a typical example of which is shown in Fig. 1i, already indicate an exceptional crystalline quality of the film surfaces. Moreover, the bulk is characterized by excellent crystallinity. Neither $\theta$-$2\theta$ X-ray diffractograms nor rocking-curve measurements allowed us to observe any difference in the crystallinity between the films and the substrates. To gain further microscopic information on the film microstructures at the atomic scale, we imaged sample cross sections by STEM. A wide-field HAADF-STEM image is shown in Fig. 4a. No crystallographic defects are visible in the film at this magnification, and the interface between the film and the substrate is not even discernible. However, as described above, the interface's location and existence were determined by profilometry. LAADF is especially sensitive to defects [74], but microscopy in this mode also failed to reveal any defects (Fig. 4b). Even imaging the interface region at higher magnification does not reveal the interface (Fig. 4c, d). Indeed, we found no microscopic difference between the single-crystal sapphire substrates, and the sapphire films that were grown under the conditions of adsorption-controlled growth (Table 2).

As the possible presence of chemical impurities is a key issue affecting the quality of homoepitaxial sapphire films [11,20], we also performed chemical analyses on these films by time-of-flight secondary ion mass spectroscopy (ToF SIMS). Figure 5 compares the corresponding results for several ionic species dependent on the thickness $d$ of (a) a sample grown by MBE and (b) a sample grown by TLE.

First, we note that the TLE film grown under adsorption-controlled growth conditions is measured to be very pure. Indeed, the film shows fewer impurities than the underlying single-crystal substrate. The MBE-grown film shows fewer $B^+$ impurities than previous reports [11]. B originates from the p-BN crucibles used for Al sources, and the adsorption-controlled growth may minimize its incorporation. The dominant impurity detected in the MBE grown film is $Ti^+$. Titanium is used for back-coating the substrates.

The main other impurities detected in both films are $Ga^+$, $Si^+$, and $Fe^+$, which are all present in the single-crystal substrate. For the MBE-grown film, a typical enrichment of these impurities at the interface and surface is observed. Interestingly, this is not the case for the TLE-grown film, where the $Ga^+$ and $Fe^+$ signals fall below the detection level within a couple of nanometers. Only the $Si^+$ signal has a small peak at the substrate film interface and does not deplete within the film. The omnipresence of $Si^+$ is well known for numerous materials [75,76]. Indeed, the high solubility of Si in sapphire over wide temperature and pressure ranges makes it difficult to avoid this impurity [77].

Furthermore, owing to their high purity, the *c*-plane sapphire films grown by adsorption control are not subject to the band gap reduction that vexes standard epitaxial sapphire films [11,20]. Figure 6 shows the corresponding ultraviolet c (UVC) transmittance spectrum of a homoepitaxial *c*-plane film grown by adsorption control along with the spectra of single crystals. The band gaps of the film and the substrates are identical.

The higher purity of the TLE-grown epitaxial film compared to the single-crystalline substrate is also revealed by ultraviolet-visible (UV-Vis) spectroscopy, see Fig. S2. This is because $F^+$ centers in sapphire cause light absorbance at 260 nm [78], meaning that the lower absorbance at this wavelength provides evidence of fewer $F^+$ centers in the films compared to the substrates. Consistent with this, the lower optical conductivity measured by ellipsometry in the films indicates that the defect density in the films is at least as low as in the annealed substrates [79], see Fig. S3.

## 4. Summary and Conclusions

We have identified the deposition conditions required to achieve adsorption-controlled homoepitaxial growth of sapphire by TLE and MBE. These conditions are summarized in Table 1.

The films are characterized by excellent structural and optical properties. Compared to standard sapphire films, the films grown by adsorption control have lower defect densities despite their very high growth temperatures. The benefits of adsorption control thus outweigh the thermodynamically driven increase in defect density with increasing temperature.

The ability to grow homoepitaxial sapphire films that surpass the structural and optical properties of sapphire single crystals opens new perspectives. For example, the growth of such homoepitaxial layers allows detrimental effects resulting from defects incorporated into the surface of sapphire single crystals to be minimized, which we expect to yield enhanced mobilities, fewer losses, and enhanced quantum decoherence times [80].

Furthermore, such films with their excellent properties can serve as functional layers in electronics, e.g., for high-power applications [10–12].

As the high quality of these films is induced by the adsorption-controlled growth mode rather than by specific properties of the Al-O material system, we anticipate that additional oxide films as well as non-oxide material systems will be obtainable in excellent quality by following the approach presented here.

Acknowledgements

The authors thank Bettina V. Lotsch and Alexander Boris for providing the infrastructures of the UV-Vis spectroscopy and of the ellipsometry, respectively.

Conflict of Interest

The authors have no conflict of interest to declare.

Table 1 Growth regimes of sapphire observed in TLE and MBE for the following parameters. TLE: $P_L \geq 250$ W, $p_{ox} \geq 1\times10^{-3}$ mbar $O_2$. MBE: $\Phi_{Al} = 2.2\times10^{15}$ atoms/cm$^2$, $p_{ox} = 7\times10^{-6}$ mbar 90 % $O_3$ and 10 % $O_2$.

| $T$(°C) | Growth | |
|---|---|---|
| RT | Amorphous | |
| 900 | Multilayer | Epitaxial & adsorption controlled |
| 1300 | Step flow | |
| 1800 | Step bunching | |

Table 2 Adsorption-controlled growth regimes for MBE and TLE identified in this work.

| | Parameter regime | Ideal range |
|---|---|---|
| TLE | $T \geq 900$ °C | 1300 °C $< T \leq$ 1800 °C |
| | $p_{ox} \geq 1\times10^{-3}$ mbar $O_2$ | |
| | $P_L \geq 250$ W | |
| MBE | 900 °C $\leq T$ | |
| | $p_{ox} = 7\times10^{-6}$ mbar 80 % $O_3$ and 20 % $O_2$ | |
| | $\Phi_{Al} = 2.2\times10^{15}$ atoms/cm$^2$ | |

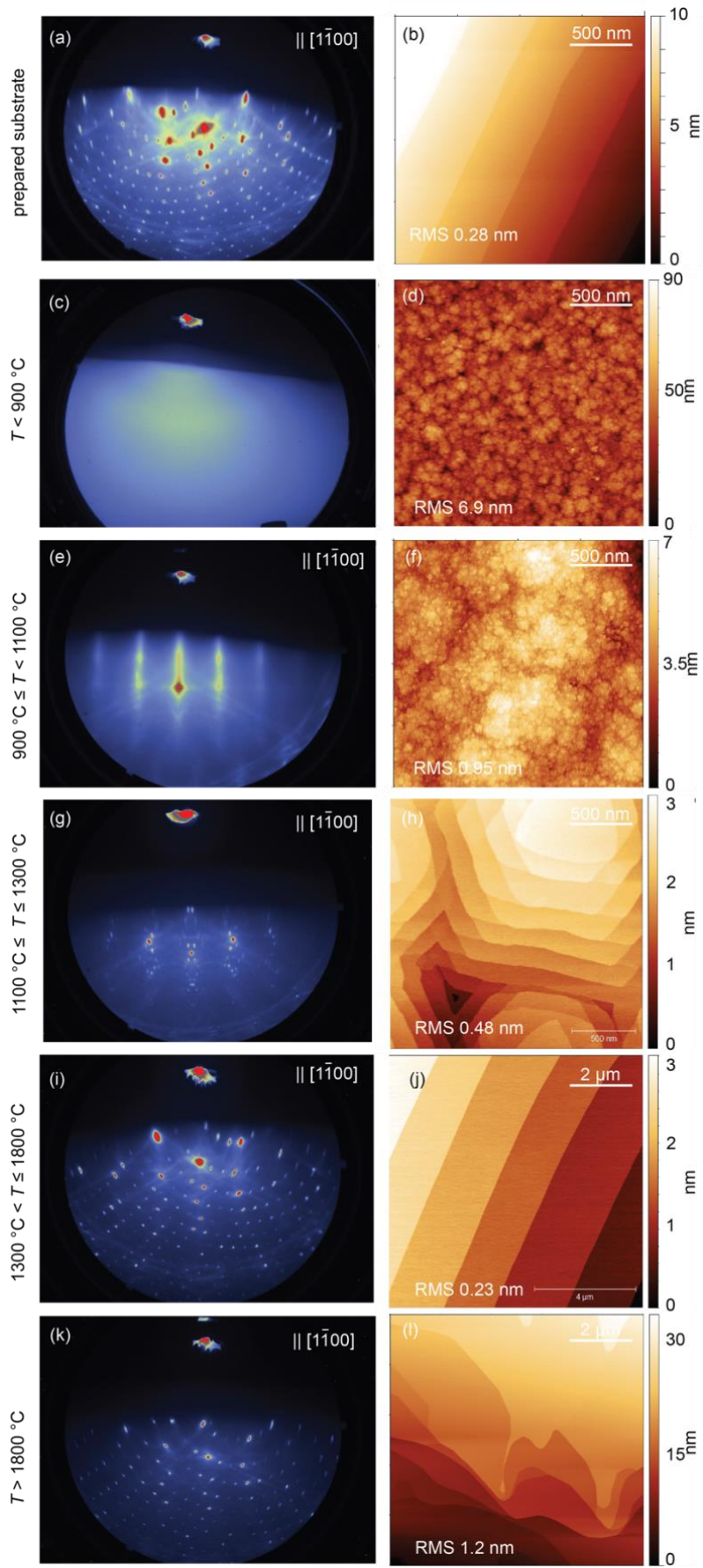

**Figure 1.** RHEED patterns (left column) and AFM topography images (right column) showing different growth regimes of Al$_2$O$_3$ on *c*-plane sapphire (TLE). The MBE-grown films show no significant difference and are therefore shown in the supplementary information.

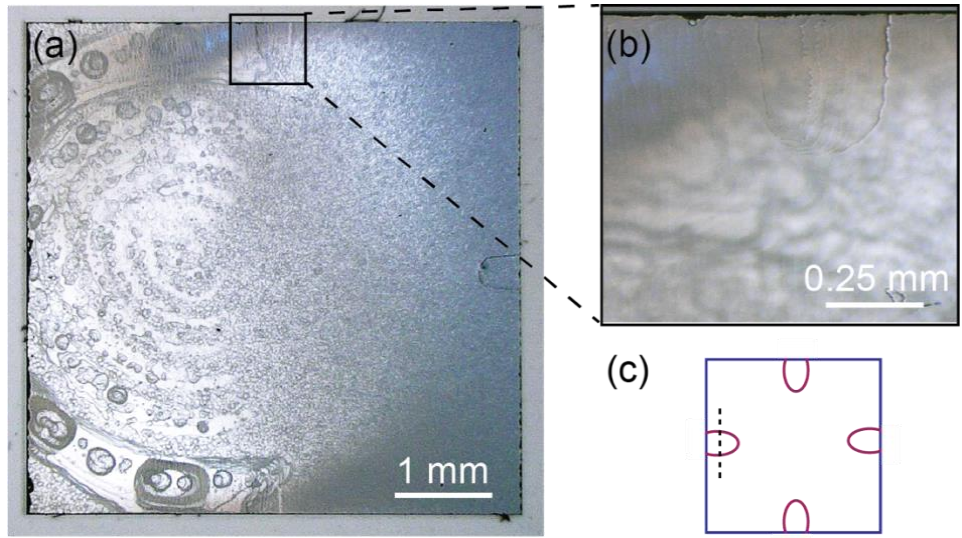

**Figure 2.** (a) Micrograph of the homoepitaxial sapphire film grown at 2000 °C (Fig. 1k–l). The melt pool and bubbles are frozen due to quenching. (b) The enlargement shows that growth is being precluded at the position of the holder. Step bunching results in steps visible to the naked eye. (c) Schematic of the sample where the areas blocked are marked in red and the dashed line exemplary shows how profilometry is carried out

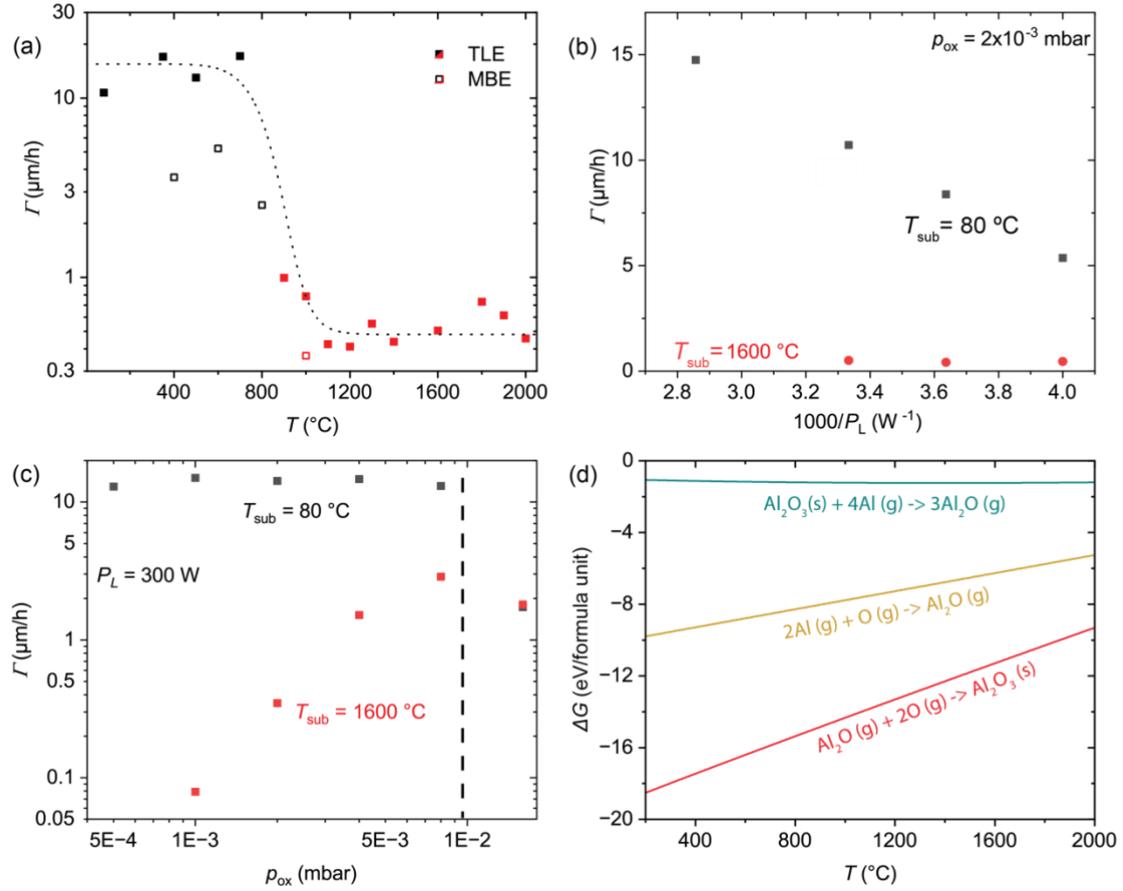

**Figure 3.** (a) Growth rate $\Gamma$ as a function of substrate temperature $T$ with open symbols marking MBE growth and closed symbols marking TLE growth. Black represents amorphous films and red represents crystalline films. The laser power $P_L$ is 300 W and the oxygen background pressure $p_{ox}$ is $2\times10^{-3}$ mbar for the TLE grown films, the Al-flux is $2.2\times10^{15}$ atoms/cm² at $p_{ox} = 7\times10^{-6}$ mbar for MBE. The dashed curve represents a model fit. (b) TLE $\Gamma$ as a function of $1000/P_L$ and (c) a function of $p_{ox}$ with the dashed line marking the pressure for which the mean free path is lower than the distance between source and substrate for an unheated substrate (black) and $T = 1600$ °C (red). (d) Gibbs free energy ($\Delta G$) as a function of $T$ for the reactions expected in this parameter regime.

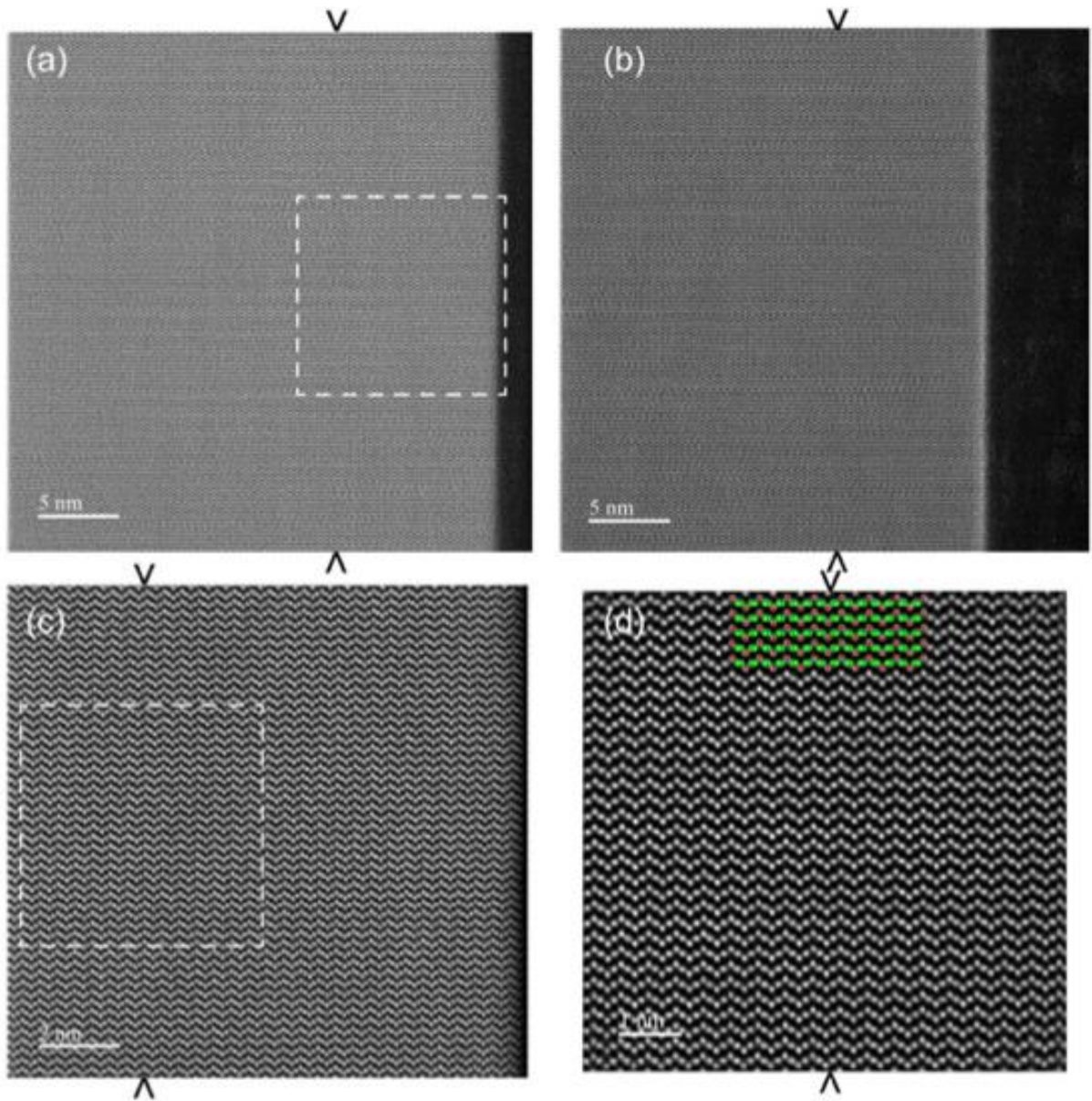

**Figure 4.** (a) ADF–STEM image showing the long-range high quality of the grown sapphire film and its interface to the substrate, the position of which is marked by V symbols. (b) Same area imaged in LAADF, which is more sensitive to strain and non-stoichiometry, neither of which are observed. (c) Enlargement of the interface region in (a) that further shows the high crystal quality. (d) HAADF–STEM of the interface region, which remains indistinguishable. Red dots correspond to O, green dots to Al. All images are taken along the $[1\bar{1}00]$ direction with the growth direction from left to right.

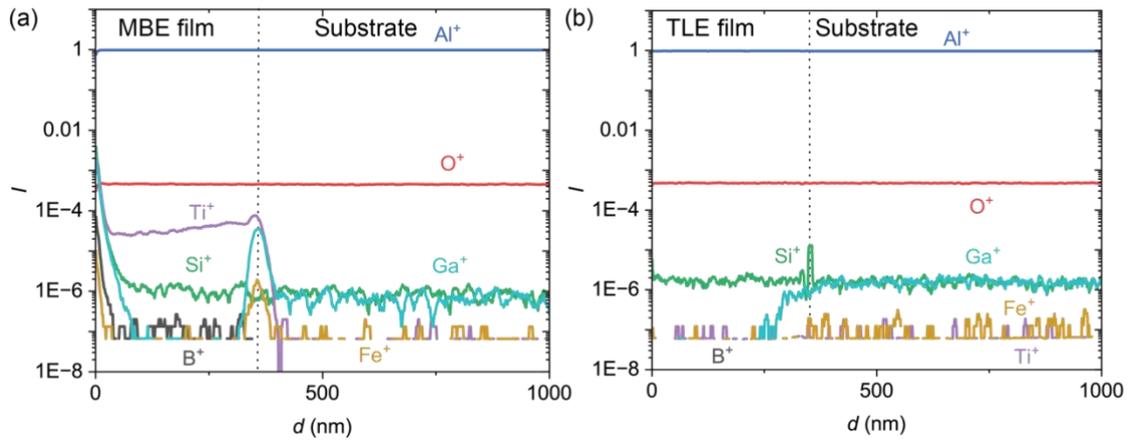

**Figure 5.** ToF SIMS measurements of a homoepitaxial sapphire films grown in the adsorption-controlled regime at substrate temperatures of (a) 900 °C in MBE and (b) 1600 °C in TLE. The position of the interface is marked by dashed lines.

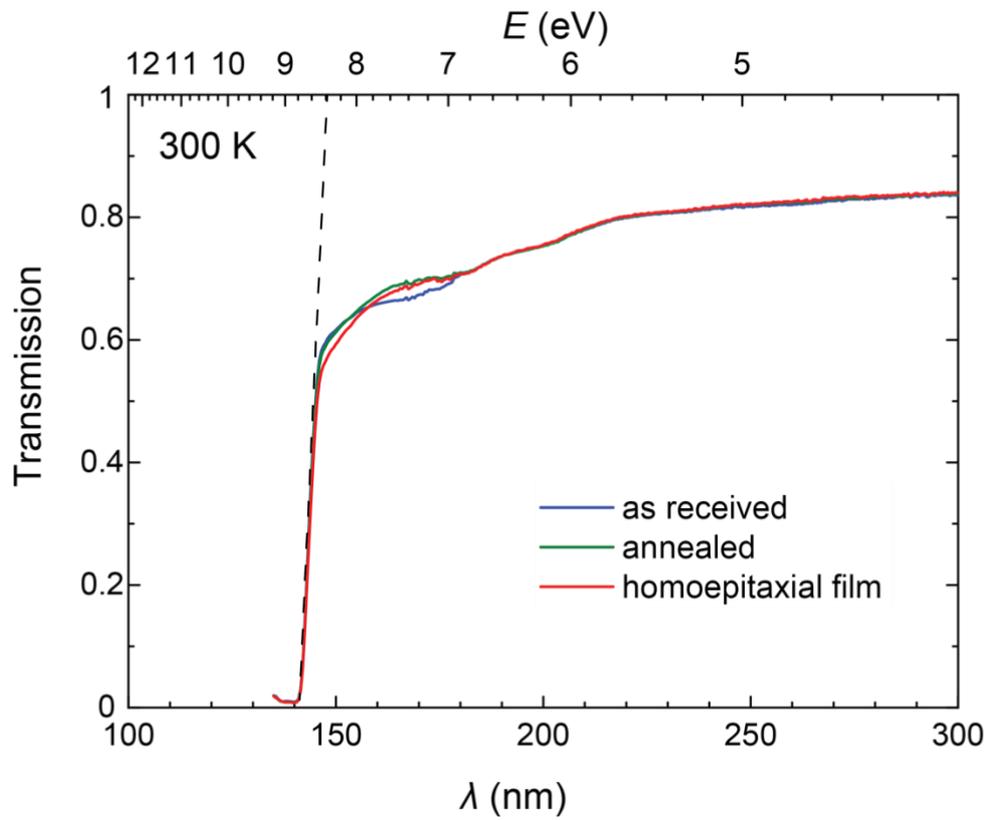

**Figure 6.** Transmittance spectra in the UVC spectral range for an as-received (blue) and annealed (green) *c*-plane sapphire single crystal, and a homoepitaxial film (red). No reduction of the band gap is observed for the film.